\documentclass{aastex62}
\usepackage[utf8]{inputenc}
\usepackage{graphicx}
\usepackage{rotating}

\title{X-rays from RS Ophiuchi's 2021 eruption: shocks in and out of ionization equilibrium}

\begin{document}
\title{X-rays from RS Ophiuchi's 2021 eruption: shocks in and out of ionization equilibrium}
\author{Nazma Islam}
\affil{Center for Space Science and Technology, University of Maryland, Baltimore County, 1000 Hilltop Circle, Baltimore, MD 21250, USA}
\affil{X-ray Astrophysics Laboratory, NASA Goddard Space Flight Center, Greenbelt, MD 20771, USA}

\author{Koji Mukai}
\affil{Center for Space Science and Technology, University of Maryland, Baltimore County, 1000 Hilltop Circle, Baltimore, MD 21250, USA}
\affil{CRESST II and X-ray Astrophysics Laboratory, NASA Goddard Space Flight Center, Greenbelt, MD 20771, USA}

\author{J.L. Sokoloski}
\affil{Columbia Astrophysics Laboratory, Columbia University, New York, NY 10027, USA}

\correspondingauthor{Nazma Islam}
\email{nislam@umbc.edu}

\begin{abstract}
The recurrent nova RS Ophiuchi (RS~Oph) underwent its most recent eruption on 8 August 2021 and became the first nova to produce both detectable GeV and TeV emission. We used extensive X-ray monitoring with the Neutron Star Interior Composition Explorer Mission (NICER) to model the X-ray spectrum and probe the shock conditions throughout the 2021 eruption. The rapidly evolving NICER spectra consisted of both line and continuum emission that could not be accounted for using a single-temperature collisional equilibrium plasma model with an absorber that fully covered the source. We successfully modelled the NICER spectrum as a non-equilibrium ionization collisional plasma with partial-covering absorption. The temperature of the the non-equilibrium plasma show a peak on Day 5 with a $kT$ of approximately 24~keV. The increase in temperature during the first five days could have been due to increasing contribution to the X-ray emission from material behind fast polar shocks or a decrease is the amount of energy being drained from shocks into particle acceleration during that time period. The absorption showed a change from fully covering the source to having a covering fraction of roughly 0.4, suggesting a geometrical evolution of the shock region within the complex global distribution of the circumstellar material. These findings show the evidence of the ejecta interacting with some dense equatorial shell initially and with less dense material in the bipolar regions at later times during the eruption.

\end{abstract}

\section{Introduction}
Novae are transient systems containing a white dwarf (WD) accreting matter from a non-degenerate stellar companion. These systems go into eruption when the accreted matter undergoes a thermonuclear runaway (TNR) on the surface of the WD. The TNR releases a large amount of energy, which causes the ejection of the accreted matter. If the donor star is a red giant (RG) that produces a wind or focused outflow, the ejecta from the nova can interact with this circum-binary matter \citep{warner1995,bode2008,starrfield2016}. 
\par
The recurrent nova RS~Ophuichi (RS~Oph) consists of a massive WD and a RG donor star in a binary orbit with a period of 453.6 $\pm$ 0.4 days \citep{brandi2009}. The distance estimate to RS~Oph by the Gaia Data Release 3 is 2.44 kpc \citep{bailerjones2021}; however there is some uncertainty on the Gaia estimate because of the surrounding nebula and the wobble of the long binary period. It has previously undergone nova eruption approximately every 15 years, with recorded eruptions in 1898, 1933, 1958, 1967, 1985 and 2006 and a possible additional eruption in 1907. RS~Oph began its recent eruption on 08 August 2021 as per the discovery in the optical reported in AAVSO Alert Notice 752\footnote{https://www.aavso.org/aavso-alert-notice-752}. Subsequently, the eruption was detected across the electromagnetic spectrum: in radio with the Very Large Array (VLA) Low-band Ionospheric and Transient Experiment (VLITE \citealt{peters2021}), the Jansky VLA \citep{sokolovsky2021}, the Arcminute Microkelvin Imager Large Array (AMI-LA), e-MERLIN, and MeerKAT \citep{williams2021}; in infrared with the NASA Infrared Telescope Facility \citep{woodward2021a, woodward2021b}; in optical with the Southern African Large Telescope (SALT \citealt{mikolajewska2021}), the Varese telescope \citep{munari2021} amongst many others; in X-rays with the Monitor of All Sky X-ray Image (MAXI; \citealt{shidatsu2021ATel}), Neil Gehrels Swift Observatory (Swift; \citealt{page2021a, page2022}), INTErnational Gamma-Ray Astrophysics Laboratory (INTEGRAL; \citep{ferrigno2021}), Neutron Star Interior Composition Explorer (NICER; \citealt{enoto2021}), Nuclear Spectroscopic Telescope Array (NuSTAR; \citealt{luna2021}), AstroSat \citep{rout2021}, Chandra and XMM-Newton \citep{orio2021}. It became the first nova to have detectable TeV/GeV gamma-rays, with TeV emission detected by the High Energy Stereoscopic System (HESS; \citealt{wagner2021,hess2022}), Major Atmospheric Gamma Imaging Cherenkov Telescopes (MAGIC; \citealt{acciari2022}) and GeV emission detected with Fermi-LAT \citep{cheung2021ATel,cheung2022}. 
\par
Unlike in majority of novae hosted in CVs, the donor star in RS~Oph is a RG, and the high velocity ejecta from the TNR runs into its dense stellar wind, giving rise to X-ray emission from hot, shocked gas. In this and prior eruptions, the initial X-ray emission was dominated by hard emission in the 1--10~keV energy range thought to originate from such shocks, analogous to hard X-ray emitting shocks seen in supernova remnants \citep{vink2012}. RS~Oph thus provides an excellent laboratory to study the interaction of the ejecta with the stellar winds from the donor star. After the nova ejecta expand and become optically thin to soft X-rays, the X-ray emission is generally dominated by a blackbody-like supersoft component originating in the photosphere of the nuclear shell-burning WD with $kT$ corresponding to an effective temperature of 20--100 eV. 
\par
From numerical simulations of RS~Oph in quiescence and during eruption, we expect the quiescent mass transfer to produce a ring-like equatorial density enhancement, with which the ejecta interact and produce a bipolar outflow \citep{booth2016}. The Rossi X-ray Timing Explorer (RXTE) Proportional Counter Array (PCA) observations of the 2006 eruption of RS~Oph indeed suggested that the hard X-ray emission from that event faded at a rate that was consistent with an X-ray emitting volume increasing roughly as the square of the blast wave radius, as expected for a shock moving into an equatorial ring \citep{sokoloski2006,obrien2006,orlando2009}. Radio imaging weeks to months after the beginning of the 2006 eruption revealed an asymmetric shock wave, with a bipolar jet-like outflow and  possible equatorial enhancement in the RG wind \citep{obrien2006,rupen2008,sokoloski2008}. Optical imaging of the RS~Oph nebular remnant using Hubble Space Telescope and ground-based observations showed that the remnant consisted of two distinct components -- a high density, low velocity RG wind in the equatorial regions of the binary, and an extended, high-velocity structure at the poles \citep{ribeiro2009}. Finally, extended X-ray emission observed by Chandra after the 2006 eruption was consistent with the bipolar outflows seen in the radio and optical imaging \citep{montez2022}. So, radio, optical and X-ray observations after the 2006 eruption support the picture of RS~Oph as having a density enhancement near the equatorial plane that plays a role in shaping the nova shocks and ejecta.        
\par
During the 2021 eruption of RS~Oph, NICER continuously monitored from Day 2 through Day 90 after the beginning of the eruption, with exposure times of $\geq$ 1~ks. An analysis of the NICER data using a collisional equilibrium plasma model have already been published by \cite{orio2023}. This paper presents an independent analysis in which we explore the need for emission from a non-equilibrium ionization collisional plasma to explain the observed X-ray emission. 

\section{NICER and MAXI observations and data reduction}

\begin{figure}
\centering
\begin{minipage}{.55\textwidth}
  \includegraphics[scale=0.4]{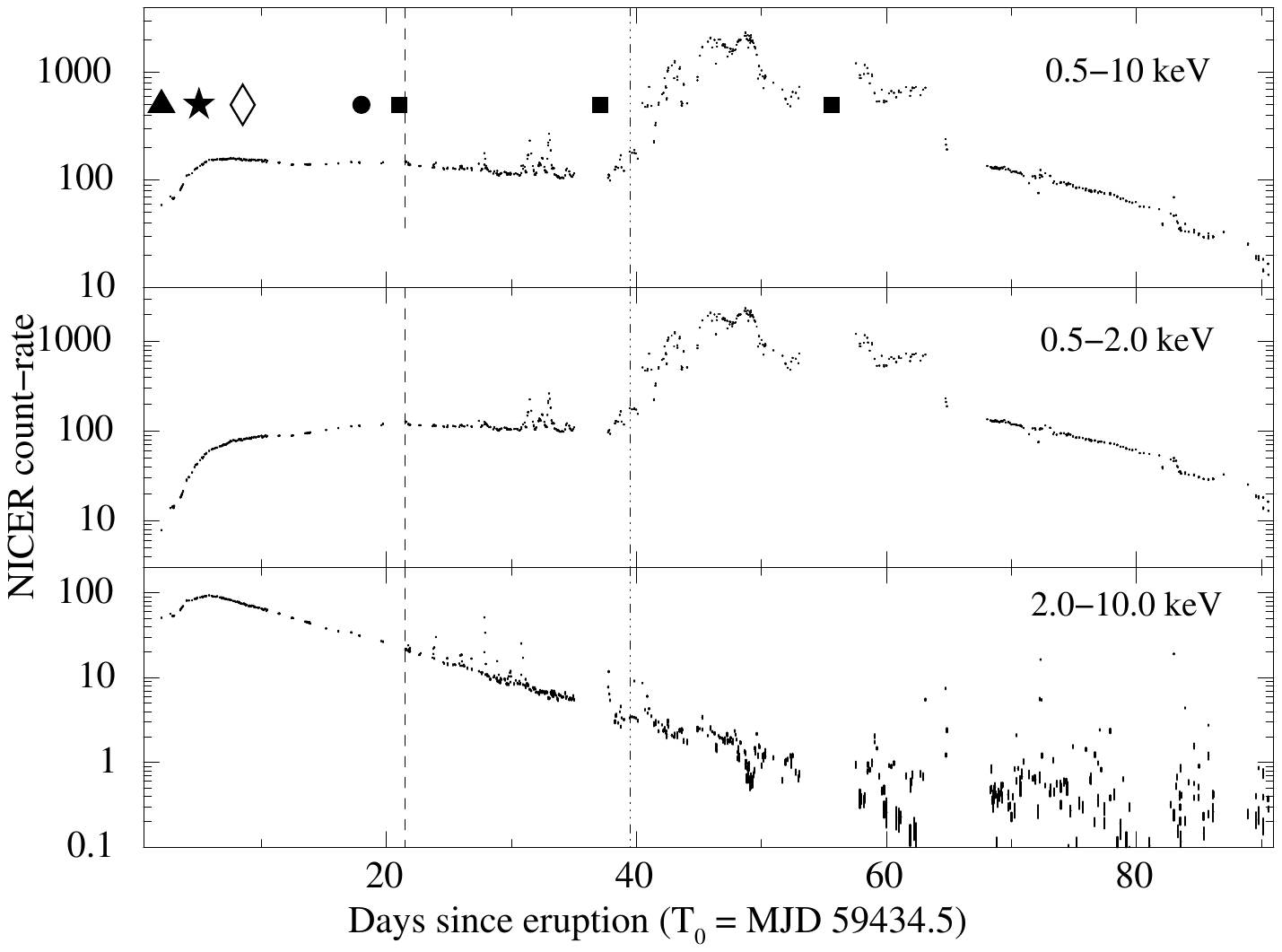}
  \caption{NICER lightcurves of RS~Oph in units of counts/sec, plotted from the beginning of the eruption (T$_{0}$ = MJD 59,434.5), with 1000 s bins. The top panel shows the lightcurve in the 0.5-10 keV band, the middle panel in the 0.5-2.0 keV band, and the bottom panel in the 2-10 keV energy band. The dashed line at Day 21 marks the beginning of the supersoft phase as reported by \cite{page2021b}. The dot-dashed line at Day 40 shows the beginning of the highly variable supersoft phase, during which the 0.5-10 keV and 0.5-2.0 keV lightcurves show rapid brightness variations by a factor of 10 or more. The triangle denotes the time of the peak of the Fermi-LAT lightcurve on Day 2, the star the time of the peak of the HESS lightcurve on Day 5, the rhombus the time of the NuSTAR observation carried out on Day 8, the circle the time of the Chandra HETG observation on Day 18 \citep{orio2022}, and the squares the times of the XMM observations on Days 21, 37 and 55 \citep{orio2022,ness2023}.}
  \label{lc}
\end{minipage}%
\hspace{10mm}
\begin{minipage}{.35\textwidth}
 \hspace{-10mm}
  \includegraphics[scale=0.3, angle=-90]{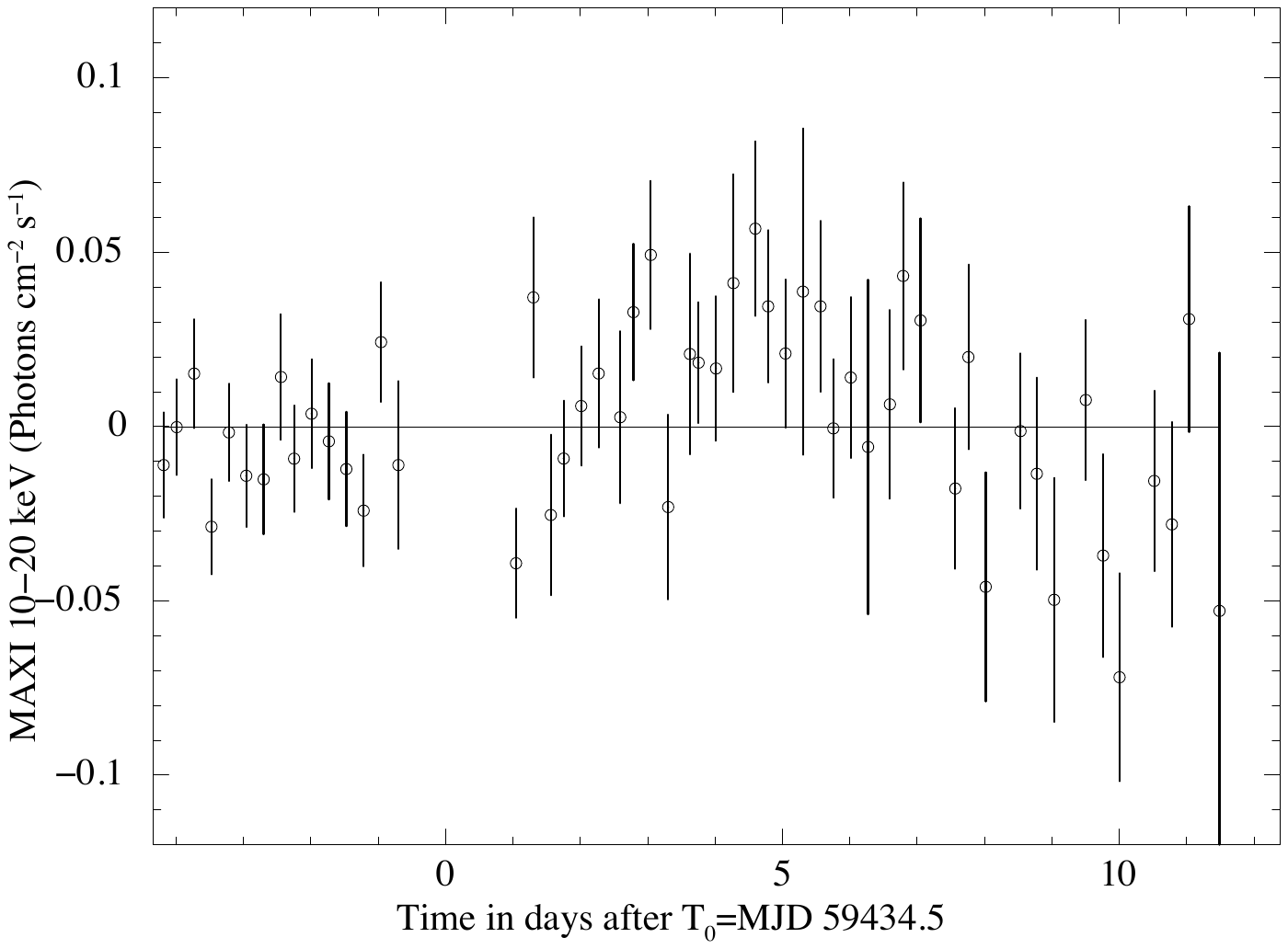}
  \caption{MAXI GSC lightcurve of RS~Oph in the 10--20 keV energy band, extracted with 5400 s bins. Consistent with the shape of the NICER lightcurves (which are at lower energies), there is a slight enhancement in the X-ray intensity around Day 4--7.}
  \label{maxi_lc}
\end{minipage}
\end{figure}

NICER is an X-ray observatory that has been operating on the International Space Station (ISS) since 2017 \citep{gendreau2016}. The main instrument, X-ray Timing Instrument (XTI), consists of 56 identical and co-aligned cameras, each containing an X-ray Concentrator (XRC; \citealt{okajima2016}) and an Si drift detector positioned in the concentrator's focal plane. During the 2021 eruption of RS~Oph, NICER monitored the progress of the eruption with almost daily exposures of between approximately 1 and 10~ks from 2021 August 10 to November 7, until RS~Oph entered the solar observing constraint. We analyzed the NICER data using {\tt nicerversion} 2022-12-16$\_$V010a, released with HEAsoft v6.31.1 and calibration files {\tt xti20221001}. The standard pipeline processing was used to carry out screening and filtering of data and apply the latest calibration files using {\tt nicerl2}. Using NICERDAS tasks {\tt nicerl3-spect}, we extracted the source, the background spectrum and the appropriate response matrices and ancillary files. The source spectrum was automatically rebinned with {\tt nicerl3-spect} using {\tt ftgrouppha} by the optimal binning method \citep{kaastra2016}. We carried out the spectral analysis by fitting the source spectrum and modelling the background spectrum using the SCORPEON background model.\footnote{https://heasarc.gsfc.nasa.gov/docs/nicer/analysis$\_$threads/scorpeon-xspec/} The NICER spectrum was analyzed in the energy range 0.5--10.0 keV and a systematic error of 3\%  in quadrature was added in the spectral fits to account for the possible uncertainties due to dust scattering effects around RS~Oph \citep{smith2016}. Using the NICERDAS task {\tt nicerl3-lc} with the Space Weather model for background\footnote{https://heasarc.gsfc.nasa.gov/docs/nicer/analysis$\_$threads/nicerl3-lc/}, we extracted the background subtracted lightcurves in different energy bands (0.5--10 keV, 0.5--2 keV and 2--10 keV). All times are relative to the start of the optical eruption at T$_{0}$ = MJD  59,434.5 \citep{munari2021}. 
\par
The NICER lightcurves reveal variations in the X-ray brightness throughout the eruption. Fig.~\ref{lc} show lightcurves constructed from the NICER observations in the energy bands 0.5--10 keV ($broad$), 0.5--2 keV ($soft$) and 2--10 keV ($hard$). The lightcurves in the broad, soft and hard energy-bands all show rapid increases in count-rate through Day 6 and then remain at a similar count-rate level through Day 21 in the broad and soft bands. The dashed line at MJD 59,456 (Day 21) in Fig.~\ref{lc} marks the beginning of the supersoft phase as reported by \cite{page2021b}. The count-rate in the hard band monotonically decreased after reaching its peak around Day 6. The count-rate in the soft band dominated the total X-ray emission from Day 20 onwards. The dot-dashed line at MJD 59,474 (Day 40) in Fig.~\ref{lc} indicates the beginning of the highly variable supersoft phase, during which the broad and soft lightcurves show rapid brightness variations by a factor of 10 or more, especially in the soft band. The contribution to the overall X-ray emission from the hard band was negligible after the source entered the highly variable supersoft phase and hence we carry out analysis till Day 40. Table 1 show the details of the NICER observations used in this paper.  
\par
MAXI is an all sky X-ray monitor that has been operating on the ISS since 2009 \citep{matsuoka2009}. The main instrument onboard is the Gas Slit Camera (GSC), which consists of six units of large area, position-sensitive Xenon proportional counters in the energy range 2--30 keV and has an instantaneous field of view of 1.5$^{\circ} \times$ 160$^{\circ}$. It covers 85 percent of the entire sky approximately every 92 minutes {\it i.e} during one ISS orbit. The in-orbit performance of GSC is summarized in \cite{sugizaki2011}. We extracted the lightcurve in the 10--20 keV energy band starting around the beginning of the 2021 eruption using MAXI on-demand data\footnote{http://maxi.riken.jp/mxondem/index.html} with a bin time of 5400 s. Fig.\ref{maxi_lc} shows this MAXI/GSC lightcurve. The source was outside MAXI/GSC field of view when the optical eruption was reported on MJD 59,434.5. There was an increase in the X-ray count-rate observed with MAXI GSC above 10 keV in the first seven days of the eruption, with the peak of the X-ray intensity occurring at around Day 4--7. 

\section{Results -- fits to the NICER spectra}

\begin{figure}
\centering
\begin{minipage}{.4\textwidth}
  \includegraphics[scale=0.32]{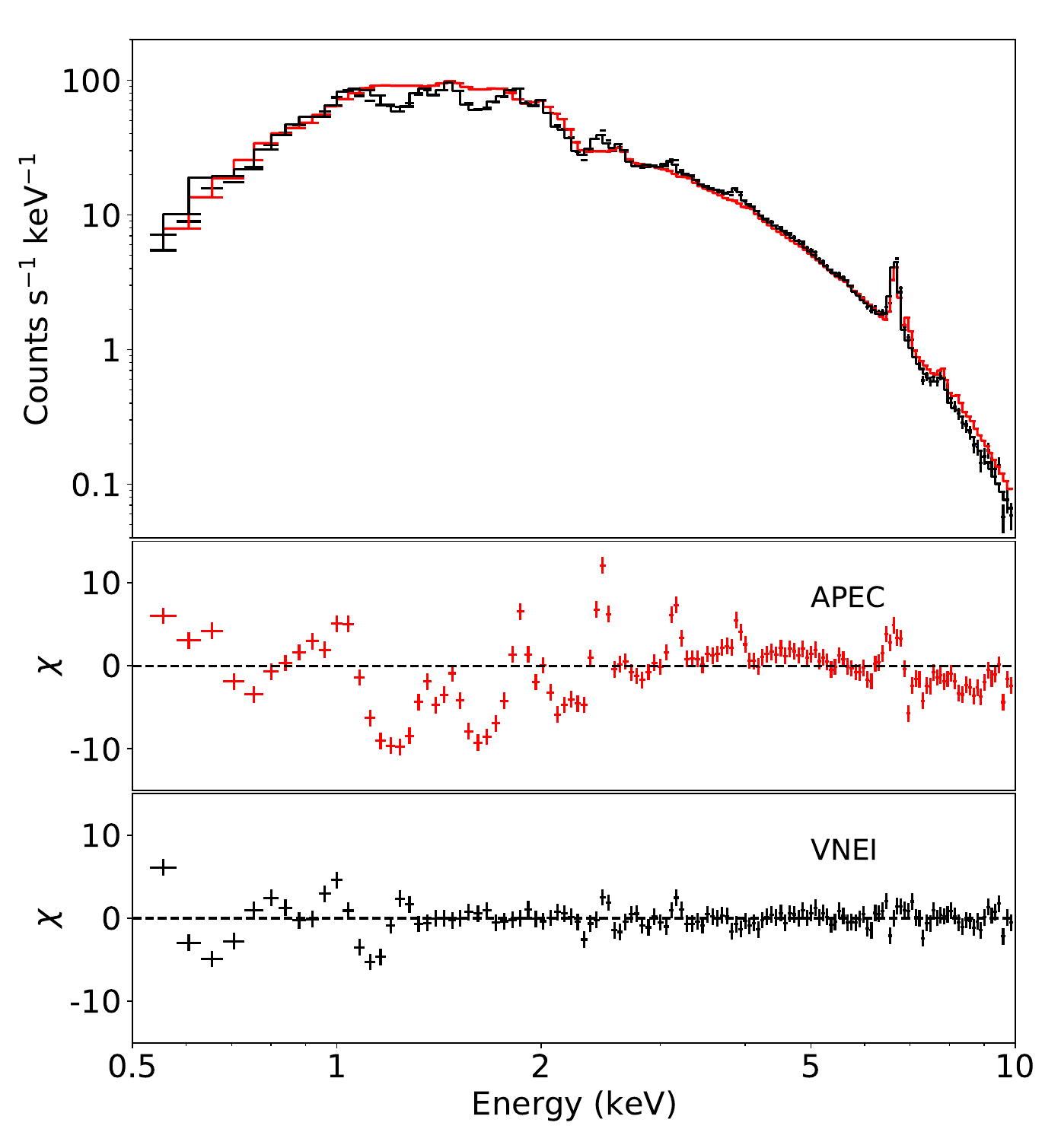}
  \caption{NICER spectrum of RS~Oph taken around Day 11. The top panel shows the NICER spectrum fitted with a single-temperature, velocity- and thermally-broadened collisional equilibrium plasma emission {\tt bvapec} model convolved with a fully covering photo-electric absorption model {\tt Tbabs} (red lines) and a model consisting of a grid of non-equilibrium {\tt vnei} components described in Section 3.2 (black line). The middle panel show the residuals for the spectral fit for the single temperature {\tt bvapec} model with fully covering photo-electric absorption. The bottom panel show the residuals for the spectral fit using the NEI grid model. The $y$ axes on both residuals plots use the same range.}
  \label{apec_spec}
\end{minipage}%
\hspace{10mm}
\begin{minipage}{.4\textwidth}
 \includegraphics[scale=0.34]{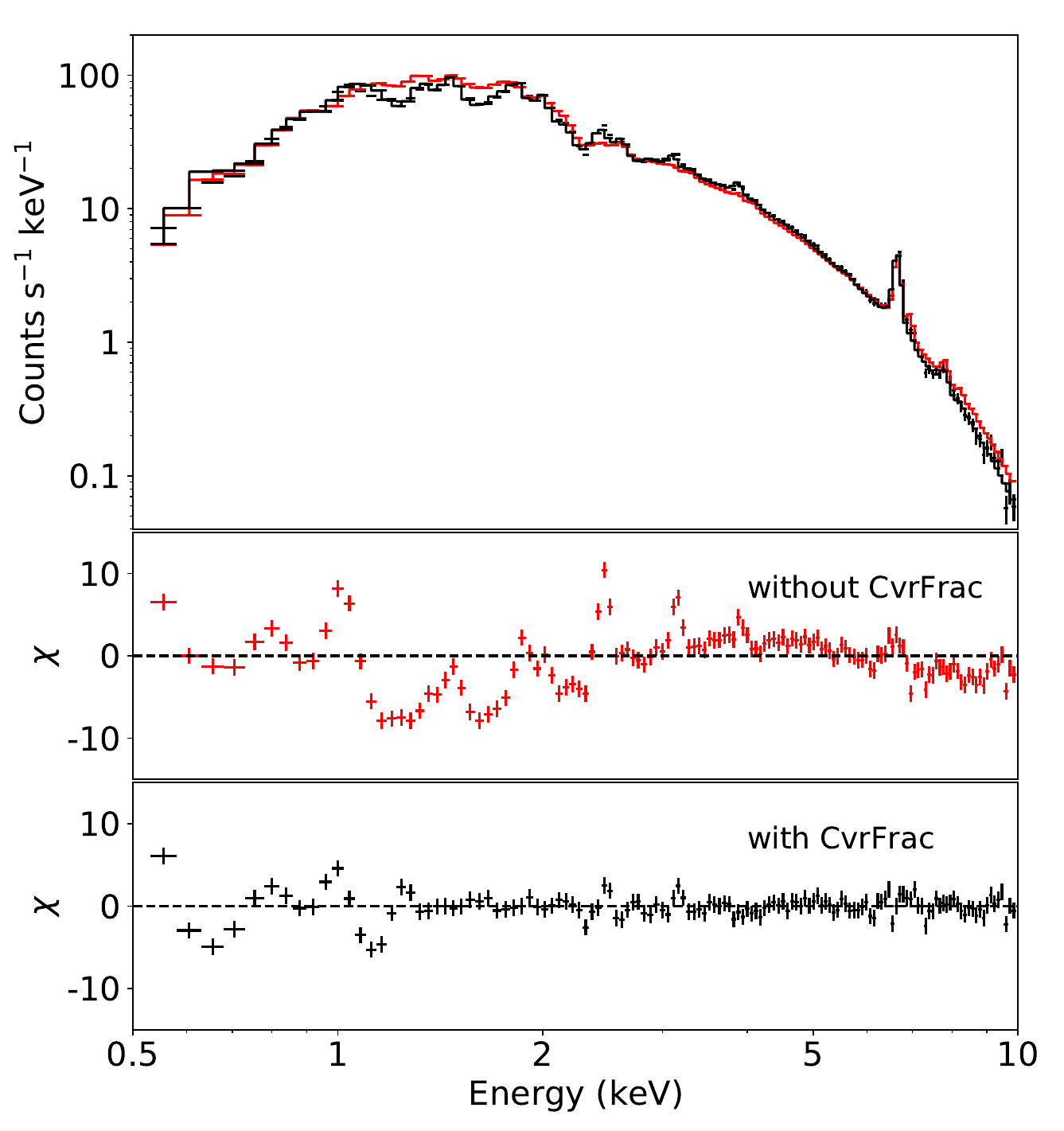}
  \caption{NICER spectrum of RS~Oph taken around Day 11. The top panel shows the NICER spectrum fitted with a model consisting of a grid of non-equilibrium {\tt vnei} components described in Section 3.2, without a partial covering absorption (red lines) and with a partial covering {\tt Tbpcf} absorption (black lines). The middle panel show the residuals for the spectral fit using the NEI grid model without a partial covering absorber. The bottom panel shows the residuals for the spectral fit using the NEI grid model with a partial covering absorber. The y axes on both residuals plots use the same range.}
  \label{pcf_spec}
\end{minipage}
\end{figure}

\begin{figure}
\centering
  \includegraphics[scale=0.4]{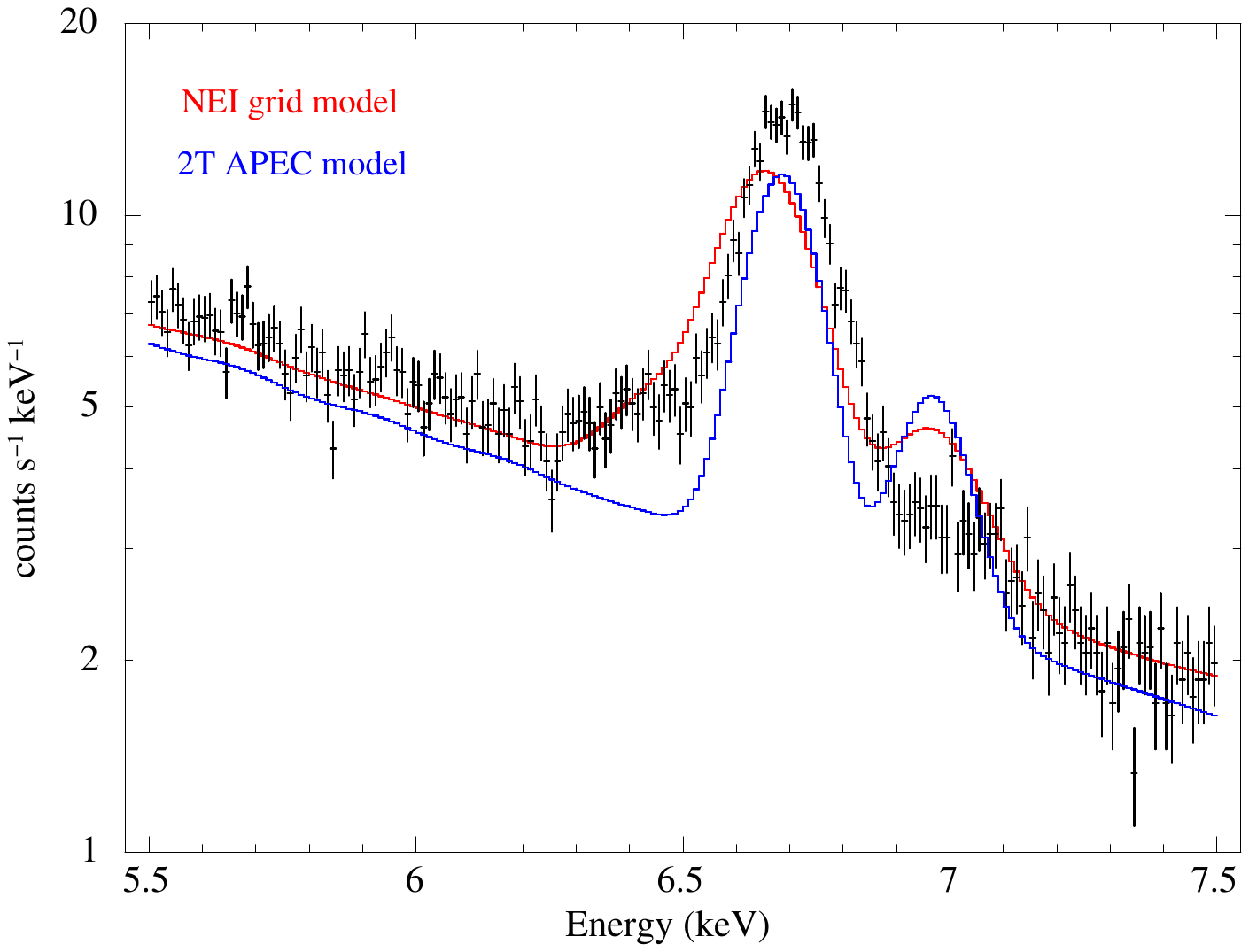}
  \caption{Fe-line region of a NICER spectrum taken around Day 4. The plot, which covers the 5.5--7.5 keV spectrum, shows He-like and H-like Fe lines. The red curve shows the fit to the Fe emission lines using the NEI grid model described in Section~3.2. The blue curve shows the fit to the Fe emission lines using a two-temperature velocity- and thermally-broadened collisional equilibrium plasma emission model {\tt bvapec} convolved with a fully covering photo-electric absorption model, {\tt Tbabs}. The NEI grid model describes the emission around 6.2--6.5 keV better than the two-temperature APEC model.}
  \label{Fe_spec}
\end{figure}

\subsection{Shortcomings of single-temperature APEC models}
The early hard X-ray emission seen during the previous 2006 eruption of RS~Oph was attributed to a shock interaction between the nova ejecta (violently ejected from the WD surface at several thousand kilometers per second) and the dense wind of the RG \citep{sokoloski2006, bode2006}. To fit the low spectral resolution RXTE data, \cite{sokoloski2006} used a model consisting of Bremsstrahlung continuum emission plus a Gaussian for the blended Fe emission lines. \cite{bode2006} used a {\tt mekal} plasma code to fit the Swift XRT data, with its typical CCD resolution but comparably modest statistical quality. With the excellent spectral sensitivity of NICER compared to RXTE PCA and Swift XRT, the NICER spectrum demand a complex spectral model.  
\par
Fig.\ref{apec_spec} shows an example fit to the NICER spectrum taken around Day 11 with a single-temperature, collisionally-ionized plasma model with variable abundances and velocity- and thermally-broadened emission lines ({\tt bvapec}); the emission component was convolved with a fully covering photo-electric absorption model {\tt Tbabs}. The residuals from the single-temperature {\tt bvapec} model fit show considerable structures around the observed K shell lines from the medium Z elements such as Ne, Mg, and Si. The continuum shape of the spectrum suggests an electron temperature of 3.9 keV, in which case these elements, if in ionization equilibrium, would be expected to be fully ionized and therefore produce only weak emission lines. However, the data show strong K shell lines from medium Z elements such as Ne, Mg, and Si, suggesting that these elements were not fully ionized. The other problem with the single-temperature collisional equilibrium plasma model with a single absorber is the presence of strong low energy residuals below roughly 1 keV, as seen in middle panel of Figure \ref{apec_spec}. The addition of one lower temperature component can help with the medium Z lines if the temperature is near the peak emissivity of these lines ($kT \approx$ 1--2 keV) or the low energy residuals if the temperature is very low ($kT < 0.5$ keV), but a single additional broad-band emission component cannot solve both problems. \cite{page2022} and \cite{orio2022} chose to add a lower temperature collisional equilibrium plasma model to account for the medium Z lines in their analysis of the Swift/XRT, Chandra HETG and XMM/RGS spectra of RS~Oph during its 2021 eruption. However, two temperature APEC models with a fully covering photo-electric absorber or a partial covering absorber fail to fit the NICER spectrum during the early days of the eruption (up to Day 5), as mentioned in \cite{orio2023}.

\begin{figure*}
\centering
\includegraphics[scale=0.55]{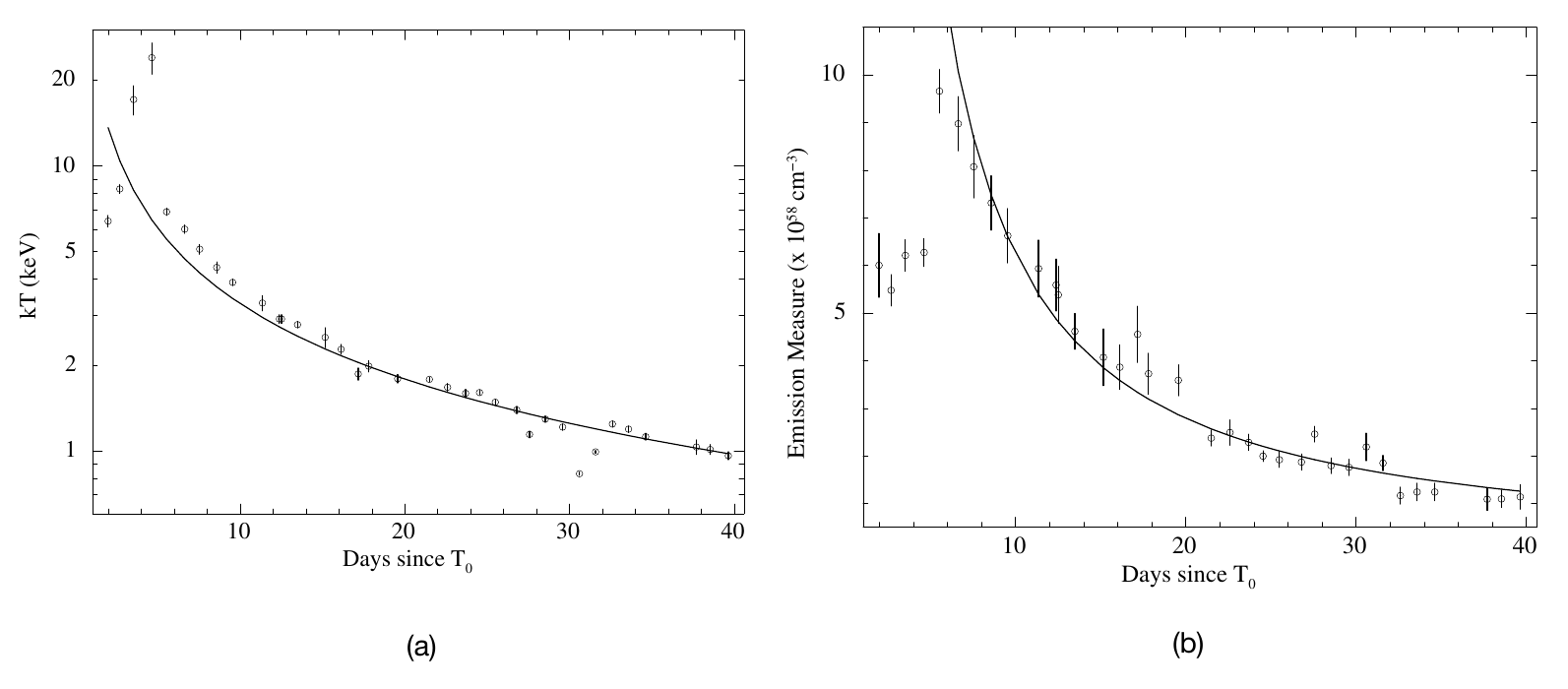}
\includegraphics[scale=0.5]{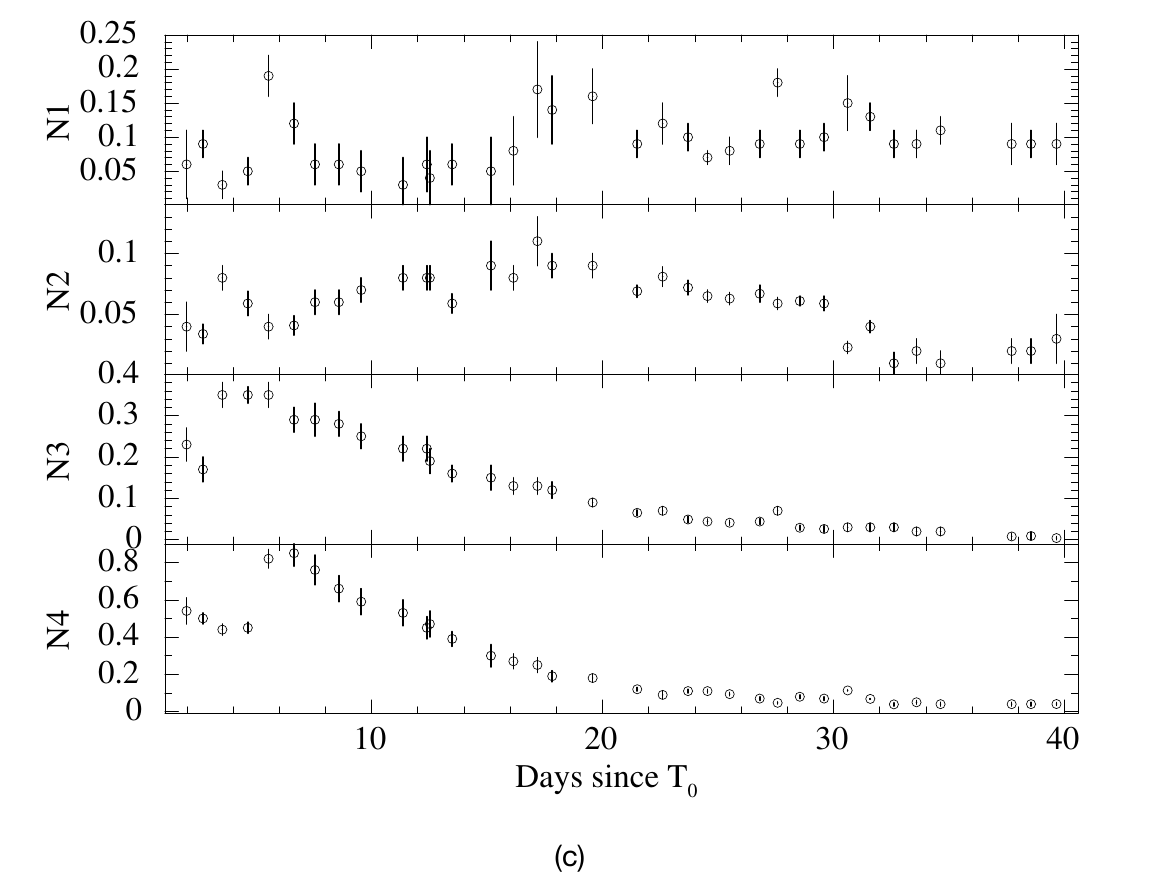}\\
\includegraphics[scale=0.55]{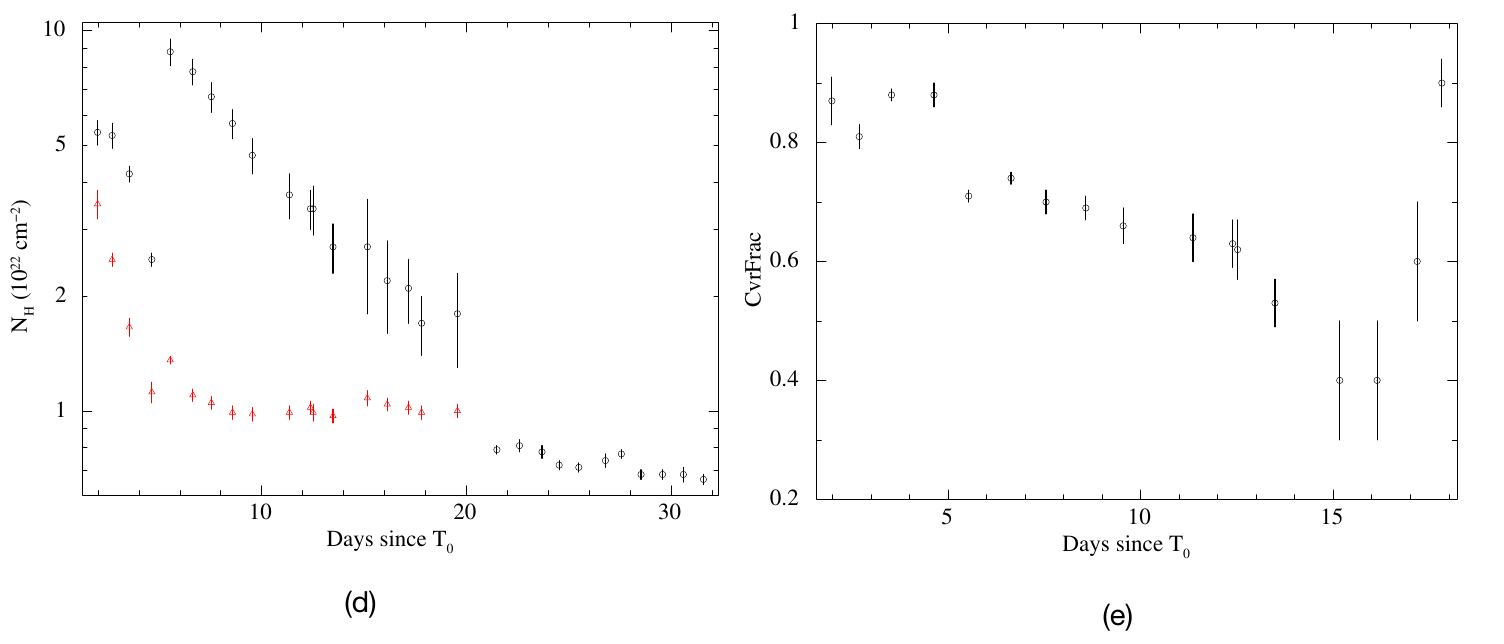}
\caption{The evolution of various spectral parameters during the first few weeks of the eruption (T$_{0}$ = MJD 59,434.5), estimated by fitting one-day averaged NICER spectrum using the NEI grid model described in Section 3.2. Panel (a): Evolution of $kT$, fitted to a power-law with an index of -0.9$\pm$0.1 (solid line). Panel (b): Evolution of EM estimated by adding the normalizations of the four {\tt vnei} components and described by a power-law with an index of -1.2$\pm$0.1. Panel (c): Normalizations of the different {\tt vnei} components: N1 is the normalization of the component with a density-weighted equilibrium timescale of $10^{9}$ s~cm$^{-3}$, N2 of that with $10^{10}$ s~cm$^{-3}$, N3 of that with $10^{11}$ s~cm$^{-3}$ and N4 of that with $10^{12}$ s~cm$^{-3}$. Panel (d): Evolution of variable absorption component from the RG wind and unshocked ejecta ($N_{\rm{Hw}}$: black circle points) and the absorption component due to the photo-electric absorption component along the line of sight ($N_{\rm{H}}$: red triangle points) in the units of $10^{22}$ cm$^{-2}$. Panel (e): Evolution of the partial covering fraction (CvrFrac). CvrFrac remained at 1.0 after Day 18, indicating that the absorber fully covered the source from Day 18 onwards.}
\label{evol}
\end{figure*}

\subsection{Non-equilibrium collisional plasma models} 
The fits to the NICER spectra using single temperature or two-temperature APEC models suffer inadequacies in modelling the emission from the K shells of medium Z elements like Ne, Mg, and Si, as well as the continuum emission below $\sim$1 keV. As previously noted by \cite{nelson2012} in the context of V407~Cyg, another nova eruption in a symbiotic binary, the existence of a second, low-temperature plasma is not the only possible explanation for the presence of medium Z lines in an X-ray spectrum.
\par
We took an alternative approach of using non-equilibrium ionization collisional plasma models. This approach is often found to be necessary in modeling the X-ray spectra of supernova remnants (for {\it e.g}: \citealt{bamba2005,badenes2006}). In astrophysical shocks, the ions that encounter a shock are initially less ionized and do not reach the equilibrium ionization level until the density-weighted timescale of the plasma, $nt$, is of order $10^{12}$ s~cm$^{-3}$, where $n$ is the electron density and $t$ is the equilibrium timescale \citep{smith2010}. The clock time required to reach ionization equilibrium is inversely proportional to the plasma density, so it is the product $nt$ that describes the degree of departure from the ionization equilibrium. If $nt$ is greater than $10^{12}$ s~cm$^{-3}$, then the plasma can be considered to be in ionization equilibrium. Values much lower than $10^{12}$ s~cm$^{-3}$ indicate that non-equilibrium ionization effects are important. If the X-ray emission originates from low density and/or freshly shocked matter, it is possible for a plasma with $kT \geq 10$~keV to nevertheless emit the K shell emission lines from Ne, Mg and Si lines.
\par
We considered several options for the inclusion of NEI effects. \cite{orio2023} modelled the NICER spectra taken on the initial days of the eruption, through Day 5, with a non-equilibrium collisional plasma model, {\tt vpshock}. While \cite{orio2023} ruled out the {\tt vpshock} model after Day 5, we believe that their analysis does not rule out NEI effects in general. {\tt vpshock} is a plane-parallel shock model that incorporates NEI physics and is appropriate for supernova remnants in the Sedov-Taylor phase \citep{borkowski2001}. In this model, the supernova remnant is assumed to be a point-like blast wave expanding into a uniform ambient medium, resulting in a linear distribution of ionization timescale versus emission measure. Given the complicated distribution of the RG wind around the WD in the RS~Oph system, it is not clear that this linear distribution would be a good approximation. We therefore constructed a model that mimics a similar single-temperature, multiple ionization timescale NEI plasma using a grid of {\tt veni} components in {\tt xspec} (vnei models are appropriate for single-temperature, single-ionization timescale NEI plasma). We take the temperature to be identical among all {\tt vnei} components, but leave the normalizations free to vary independently, thus allowing the grid to approximate an arbitrary distribution of Emission Measures (EM) as a function of ionization timescale. After several trials, we decided to implement a grid of 4 {\tt vnei} components, with the density-weighted ionization timescales of $10^{9}$, $10^{10}$, $10^{11}$, and $10^{12}$ in units of s~cm$^{-3}$, with temperatures of all the {\tt vnei} components tied to a single temperature. Furthermore, we opted to use a partial covering absorber instead of a second, lower temperature plasma to account for the low-energy residuals. We considered the NICER spectrum to be attenuated by two absorption components: a variable absorption component from the RG wind and the unshocked ejecta ($N_{\rm{Hw}}$) that only partially covers the source, modelled by {\tt Tbpcf}; and a component due to photo-electric absorption along the line of sight, modelled by {\tt Tbabs} ($N_{\rm{H}}$). This component models absorption due to the interstellar medium as well as a part of the intrinsic absorption that fully covers the emission region. After Day 20, the fits to the NICER spectrum in 0.5-10 keV energy range were unable to constrain both $N_{\rm{H}}$ and $N_{\rm{Hw}}$. Hence, we fixed $N_{\rm{H}}$ to the ISM value of 2.4$\times 10^{21}$ cm$^{-2}$ \citep{hjellming1986}, which is the minimum $N_{\rm{H}}$ expected along the line of sight to RS Oph. We used a Gaussian smoothing model {\tt gsmooth} to account for the velocity broadening of the emission lines. The full spectral model used was: \\{\tt Tbpcf}*{\tt Tbabs}*{\tt gsmooth}({\tt vnei} + {\tt vnei} + {\tt vnei} + {\tt vnei}). 
\par
We let the abundances of N, Si, Ne, Mg and Fe remain as free parameters, except for fits where these abundances could not be constrained. While Fe features are strong and allow its abundance to be estimated, Ni features are weaker and it is more difficult to constrain its abundance. However, if Fe abundance is allowed to vary while Ni abundance is fixed at the solar value, the latter may interfere with accurate determination of the former (particularly if Fe is significantly underabundant compared to the solar value). The observed [Ni/Fe] ratio in Galactic stars are found to be similar to solar abundances and these elements likely originate from same astrophysical processes \citep{eitner2023}. We link the Fe and Ni abundances in our spectral fits so that [Ni/Fe] ratio is constant at the solar ratio and the Fe abundance is estimated from the spectral fits. After Day 18, the both Ni and Fe abundances are fixed to the solar values. The number of free parameters for the above {\tt vnei} grid model is about 12 (assuming only two elemental abundances are kept free in any spectral fits), which is not significantly different from 11 and 15 free parameters in two temperature {\tt bvapec} model fits to the NICER and Chandra HETG data respectively \citep{orio2021, orio2022} and less than the 17 free parameters in three temperature {\tt bvapec} model fits to the XMM RGS spectra \citep{orio2022}.
\par
The bottom panel of Fig.\ref{pcf_spec} shows the residuals to the fit for an example NICER spectrum around Day 11, with the above NEI grid with a partial covering absorber model. Compared to the fits using a single temperature, fully covering APEC model (middle panel of Fig.\ref{apec_spec}), there is an improvement in the residuals at low energies and around the emission lines of Ne, Si, Mg and Fe. The middle panel of Figure \ref{pcf_spec} show the residuals for the fit to a NICER spectrum with the above NEI grid model without a partial covering absorption component. We see strong residuals at energies below about 2~keV, which improves with the use of a partial covering absorption model.
\par
Finally, to retain our focus on X-rays from the shocked plasma in RS~Oph rather than emission from the WD surface, we only modelled the NICER spectrum above 1.5~keV once the flux in the lower energies started to increase due to the emerging supersoft emission around Day 32. For NICER spectrum before Day 32, we used the above NEI grid model in the full energy range of 0.5--10.0~keV. After Day 32, the contribution of the flux in the soft X-ray band 0.5-2.0 keV increased due to the emerging soft emission (and soft X-ray flares as seen in Fig.\ref{lc}). The NEI grid model with the partial covering absorber became insufficient in modeling the NICER spectrum -- below 1.5 keV, there are strong residuals that would require additional complex spectral modelling with WD atmosphere models as demonstrated in \cite{page2022}, which is beyond the scope of this work. Hence, we fit the NICER spectrum in the energy-range 1.5--10 keV after Day 32. The spectral model we used was: \\
{\tt Tbabs}*{\tt gsmooth}({\tt vnei} + {\tt vnei} + {\tt vnei} + {\tt vnei}).
\par
The NICER spectrum above 1.5 keV is not sensitive to variations in the absorption, therefore we fixed $N_{\rm H}$ to the ISM value of 2.4$\times 10^{21}$ cm$^{-2}$. We also fix the abundance of N at solar since the N emission lines are below 1.5 keV and cannot be constrained by the NEI grid model for fits above 1.5 keV. After Day 40, the source entered a highly variable supersoft phase (Fig.\ref{lc}), during which the contribution of the hard X-rays to the spectrum above 1.5 keV was not significant enough to allow for NEI grid model fitting. Therefore we report results only up to the beginning of the highly variable supersoft phase. Table 2 reports the results of the spectral fits to the NICER spectrum using the above spectral model, along with error bars reported as 90\% confidence limits.
\cite{orio2023} reported on MJD 59,442 an unabsorbed X-ray flux in 0.2--10 keV energy-band of 2.2 $\times 10^{-9}$ ergs s$^{-1}$ cm$^{-2}$ using the 2T {\tt bvapec} model and 7.6 $\times 10^{-9}$ ergs s$^{-1}$ cm$^{-2}$ for the NEI {\tt vpshock} model. For a distance of 2.4 kpc, this corresponds to an unabsorbed X-ray luminosity of 1.5 $\times 10^{36}$ ergs s$^{-1}$ and 5.3 $\times 10^{36}$ ergs s$^{-1}$ for 2T {\tt bvapec} and {\tt vpshock} model respectively. For the same date, we report an unabsorbed X-ray flux in 0.2--10 keV (extrapolated from the 0.5--10 keV spectral fit) of 3.6 $\times 10^{-9}$ ergs s$^{-1}$ cm$^{-2}$. For a distance of 2.4 kpc, the unabsorbed X-ray luminosity is 2.5 $\times 10^{36}$ ergs s$^{-1}$, which is not as high as the previously reported unabsorbed X-ray luminosity using the {\tt vpshock} model.
\par
Fig.\ref{Fe_spec} further demonstrates the need for NEI models. It shows an a portion of a NICER spectrum from the early days of the eruption around Day 4, near the He- and H-like Fe lines at around 6.7 keV and 6.9 keV, respectively. The temperature determined from the continuum shape of the spectrum of the plasma was 17 keV, so if the plasma was in ionization equilibrium, we would expect the Fe atoms to be fully ionized. We fit the NICER spectrum around the H- and He- like Fe lines in two ways -- using the NEI grid model described above, and using a two-temperature APEC model. As seen from Fig.~\ref{Fe_spec}, both the models describe qualitatively the peak around 6.7 keV and predict a higher flux than observed around the 6.97 keV line. However, the NEI grid model is found to fit the rising part of the 6.7 keV line (6.1--6.4 keV) better than the two-temperature APEC model. The reason we were unable to fit the He-like and H-like Fe lines is likely our chosen spacing of ionization timescales ($10^{9}$, $10^{10}$, $10^{11}$, and $10^{12}$ in units of s~cm$^{-3}$). The Fe K complex in an NEI model with $kT$=17 keV and $nt$=10$^{12}$ s~cm$^{-3}$ is dominated by the H-like Fe line at 6.97 keV with a much weaker He-like line at 6.7 keV, whereas a $kT$=17 keV, $nt$=10$^{11}$ s~cm$^{-3}$ model is dominated by the He-like Fe line with very little H-like Fe line. The lower $nt$ models contribute the line flux at 6.4 keV. The weakness of the 6.97 keV line in the data shows that there is little near-equilibrium ($nt=10^{12}$ s~cm$^{-3}$), $kT$=17 keV plasma on Day 4, but some plasma above the next highest $nt$ grid point (i.e., $nt > 10^{11}$ s~cm$^{-3}$) may well be present. However, this constraint is derived only from the 3 Fe line fluxes: this is insufficient for us to be able to derive a quantitative model of the true EM distribution, since we might need to adjust the spacing of the $nt$ grid.

\subsection{Evolution of spectral parameters}
Fig.\ref{evol} shows the evolution of the spectral parameters estimated by fitting the above spectral model to the one-day averaged NICER spectrum. Panel (a) shows the evolution of non-equilibrium plasma temperature $kT$ (in keV). The temperature initially rose, through Day 5, when it reached a peak temperature of 24 keV. A similar enhancement in the X-ray intensities around Day 4--7 is marginally apparent in the MAXI GSC lightcurve in the 10--20 keV energy band (Figure~\ref{maxi_lc}), suggesting that the plasma temperature peak on Day 5 was real and not an artifact of spectral modelling. In contrast, the dip in measured $kT$ on Day 30-32 is most likely an artifact of the change in energy band of spectral fitting from 0.5--10 keV to 1.5--10 keV, as discussed in Section 3.2. The temperature evolution after Day 5 is reasonably well described by a power-law with an index of -0.9$\pm$0.1. This index is not too different to that of -0.7 expected for the temperature evolution of plasma heated by a decelerating blast wave moving into a spherically symmetric stellar wind, even though the circumstellar distribution around RS~Oph is known to be atleast somewhat asymmetrical \citep{sokoloski2006}.
\par
The Emission Measure (EM) peaked at around Day 6 (a day after the peak in $kT$) and then monotonically decreased as the eruption progressed. EM can be approximated as the sum of the normalizations of the four {\tt vnei} components used in the spectral fitting. Panel (b) of Fig.~\ref{evol} shows the evolution of the EM. After Day 6, the decrease in the EM is reasonably well described by a power-law with an index of -1.2$\pm$0.1. Panel (c) of Fig.~\ref{evol} shows the evolution of the individual normalizations of the four {\tt vnei} components: N1 is the normalization of the vnei component with a density-weighted equilibrium timescale of $10^{9}$ s~cm$^{-3}$, N2 of the one with $10^{10}$ s~cm$^{-3}$, N3 of the one with $10^{11}$ s~cm$^{-3}$ and N4 of the one with $10^{12}$ s~cm$^{-3}$. From the earliest days of the eruption through Day 17, the values of N3 and N4 were larger than N1 and N2 and then showed a monotonic decline as the eruption progressed. The values of N1 and N2 remained fairly constant throughout the eruption. After Day 18, the values of N1 and N2 started to dominate over the values of N3 and N4. 
\par
Panels (d) and (e) of Fig.~\ref{evol} show the time evolution of the column density of the intrinsic, variable absorption from the RG wind and the unshocked ejecta ($N_{\rm{Hw}}$), absorption component along the line of sight ($N_{\rm{H}}$) as well as the degree to which that absorber covered the source of X-ray emission (CvrFrac), respectively. During the first five days of the eruption, $N_{\rm{Hw}}$ dropped by about a factor of two. It then jumped up by roughly a factor of four around Day 6 and then monotonically decreased through Day 32, after which we fixed it to the ISM value. The covering fraction (CvrFrac), declined between the start of the eruption and around Day 16, when it reached a minimum value of 0.4. After Day 18, the spectra were consistent with containing only a fully covering absorber. 
\par
We kept the abundances of the NEI grid model fixed to the solar abundance, except for N, Ne, Mg, Si, Fe and Ni. The abundances of Fe and Ni were tied to each other and spanned a range of 0.9-1.5 Z$_{\odot}$. Table 2 reports the abundances obtained with the NEI grid model. The abundances of Ne were 0.9-1.0 Z$_{\odot}$, Mg were 0.6-2.1 Z$_{\odot}$ and Si were 0.7-1.7 Z$_{\odot}$. We allowed the abundance of N to vary after the source entered the supersoft state on approximately Day 20 and by around Day 32, it had reached a value of higher than 50 Z$_{\odot}$. After Day 32, the spectral fits were carried out in 1.5--10 keV and hence we could not constrain the abundance of N with the spectral fits and was fixed at 1.0. The fact that our fits do not require large deviation from solar abundances, except for N, is reassuring. The large overabundance of N implies the presence of CNO-processed material, most likely created during the thermonuclear runaway \citep{sokolovsky2020}. 

\section{Discussion}

\subsection{Evolution of post-shock X-ray emission in the context of a two-zone model}
 The evolution of the normalizations for the four {\tt vnei} components in the NEI grid model suggests that the distribution of circum-binary material deviates significantly from the distribution expected by a spherically symmetric wind from the RG. The X-ray emission during the first several weeks of the RS~Oph eruption (before Day 40) most likely arose from plasma that was shock-heated as the ejecta ran into circumstellar and circum-binary material. For any plasmas with $kT$ above a few keV, cooling is dominated by Bremsstrahlung process and the cooling time is always $\sim$1000 times longer than the time it takes plasma to reach ionization equilibrium. At a given point in time, X-ray emitting plasma with a higher density-weighted ionization timescale is closer to ionization equilibrium -- and likely more dense -- than plasma with a lower density-weighted ionization timescale. So, the evolution of the four {\tt vnei} component normalizations provides information about the density distribution through which the shocks propagated. For shocks moving into a spherically symmetric wind with a monotonically decreasing density profile, we might expect the N1 and N2 normalizations for emission from lower-density plasma to initially be low, with the N3 and N4 normalizations for emission from higher-density plasma initially high. After the shocks move into a lower density plasma, N1 and N2 increases as N3 and N4 drop. We observed the expected behavior for N3 and N4, suggesting that some shocks did indeed move through material with high density near the binary and a strongly decreasing density profile. The importance of high ionization timescale components between Day $\sim$6 and $\sim$17 (N3, N4) suggests near ionization equilibrium plasma dominated the X-ray emission at this epoch. That N1 and N2 were fairly constant throughout the first 40 days of the eruption, however, hints at some shocks propagating through lower density regions from the earliest days of the eruption. Such evolution of the N1, N2, N3, and N4 {\tt vnei} normalizations is consistent with a RG wind distribution in RS~Oph that is concentrated near the orbital plane, as suggested by \cite{booth2016}. Deviation from a spherical geometry was also suggested as a possible interpretation of the rapid decline in X-ray flux by \cite{sokoloski2006}. 
\par 
To explain the evolution of the emission components with different density-weighted ionization timescales, we envision a simple two-zone picture in which spherically symmetric nova ejecta collided with a dense equatorial ring -- one of the two zones -- and low density material in roughly conical, polar regions -- the other zone. In fact, observations do indicate that the circum-binary region consists of dense equatorial structure and less dense, bipolar regions. In particular, for over five years after the 2006 eruption, bipolar X-ray emitting flows appear to have maintained a tangential velocity of 4600 km\,s$^{-1}$,  which requires the presence of virtually un-decelerated ejecta in the polar directions \citep{montez2022}. This suggests that, in the polar directions, the density is considerably lower than what a spherically symmetric toy model would suggest. Because X-ray emission is proportional to the square of the density, we expect the early X-ray emission to have been likely dominated by the shock driven into the equatorial ring. Therefore, even if the two zones subtended similar solid angles, the X-ray luminosities of these two zones are heavily weighted towards the denser region. When the ejecta interacted with the densest regions of the RG wind, near the orbital plane, higher density-weighted ionization timescales of 10$^{11-12}$ s~cm$^{-3}$ were dominant, and hence the plasma producing most of the early emission reached equilibrium quickly. After Day 18, when the shocks from ejecta interacting with wind in the less dense bipolar regions started to become more important, the contribution to the EM from the regions with higher density-weighted ionization timescales of 10$^{11-12}$ s~cm$^{-3}$ decreased, and the contribution from the regions with lower density-weighted ionization timescales of 10$^{9-10}$ s~cm$^{-3}$ started dominating (Figure \ref{evol}c). EM decreased as the ejecta expanded into the less dense environment (Fig. \ref{evol}b).
\par
The basic two-zone picture supported by X-ray monitoring of RS~Oph may also explain a feature of the very high energy (VHE) emission from RS~Oph in eruption. The GeV Fermi-LAT lightcurve peaked on the Day 2 after the start of the optical eruption, and the TeV HESS lightcurve peaked on Day 5. \cite{cheung2022} modelled this delay in the peak of the observed TeV emission compared to Fermi GeV emission as the consequence of different particle acceleration timescales in a single-zone shock by, whereas \cite{diesing2023} modelled the delay with multiple shocks in the ejecta interacting with the asymmetric environment around the WD. Although our X-ray observations do not distinguish between these two models for the GeV and TeV emission, they do support the idea that distinct zones of the circum-binary environment gave rise to shocks with different properties. 

\subsection{Origin of the observed temperature and Emission Measure evolution}
Fig. \ref{lc} shows that the NICER count-rate in broad, soft and hard energy-bands increased through Day 6 and then remain at a similar count-rate level in broad and soft energy-bands up to Day 21, whereas the count-rate in the hard band montonically decreased after Day 6. An initial increase in the temperature through Day 5, when $kT \sim$24 keV, is seen in Figure \ref{evol}(a). It may be that a single shock dominates during these early times, whose energy goes into thermal X-rays and particle acceleration, with the efficiency of particle acceleration declining \citep{cheung2022}, raising the X-ray temperature as it does so. Alternatively, there may be fast and slow shocks at these early times, as \cite{diesing2023} argued to explain the GeV and TeV gamma-ray light curves, with the power in the slow shock peaking early and then the power in the fast shock peaking around Day 5. The continued cooling when the non-equilibrium component became dominant requires an explanation other than radiative cooling. Deceleration of the ejecta is unlikely in the low density, polar directions, while it is quite plausible in the equatorial region. An explanation for the evolution of the emission behind the polar shocks is explored in the next paragraph.
\par
So far, we have focused the possibility of non-equilibrium ionization effects. Another effect is possible for even younger and/or lower density shocks: the electron temperature may be considerably lower than the ion temperature (see {\it e.g.}, \citealt{masai1984} and \citealt{raymond2023}). Moreover, it is the electron temperature that we derive through X-ray spectral fitting, which, in such a case, cannot be used to derive the shock velocity. The initial post-shock temperature of electrons, in the absence of Coulomb collisions or other heating processes, would be of order m$_{e}$/m$_{p}$ lower than the shock temperature, so the emission of X-rays in RS~Oph proves that the electrons are heated.  In the context of supernova remnants, T$_{e}$/T$_{i}$ $\sim$ 0.1 appears consistent with observations \citep{matsuda2022}, and interpreted as due to collisionless electron heating. We speculate that, at late times during the NICER monitoring campaign, the timescale for post-shock Coulomb heating became longer and longer, as the ejecta density further decreased. In such a situation, it is plausible for us to observe lower T$_{e}$ as X-ray emission is dominated by materials in the earlier phases of the gradual Coulomb heating. \citet{montez2022} observed a constant temperature ($kT \sim $0.2 keV), freely-expanding X-ray emission region about $\sim$5 years after the 2006 eruption. This may indicate a $kT_{\rm{shock}} \sim$ 20 keV (expected for a strong shock given the inferred expansion velocity of the order of 4500 km/s) in which T$_{e}$/T$_{i} \sim$ 0.01.  At earlier times, the observed electron temperature might represent an electron population which is gradually heated up towards the shock temperature, with timescales that depends on the local density.
\par
The EM plot in Figure \ref{evol}(b) show a sudden increase in the value around Day 6. A likely scenario explaining the sudden jump in the EM could be of the ejecta interacting with a dense clump of matter around the equatorial ring, which are predicted by the hydrodynamical simulations of \cite{booth2016}. \cite{luna2020} reported a brightening in RS~Oph in 2017 in V band using AAVSO lightcurves, presumably implying that the accretion rate through the disk increased. We speculate that this may have been caused by an episode of enhanced mass loss from the RG and that this same enhancement could have produced an unaccreted clump of matter outside the binary that slowly (with a velocity of the order of 10 km/s) expanded. This previously ejected clump will be at a reasonable distance from the binary for the 2021 nova ejecta to plough into this region approximately 5 days after the start of the eruption.
\par
The actual circum-binary density structure around RS~Oph is certainly even more complex than our two-zone scenario. Any density diagnostics from observed X-ray emission may apply to the densest regions, and may or may not provide constraints on the density of the polar cones. \cite{sokoloski2006} interpreted the evolution of X-ray spectrum of RS~Oph during the 2006 eruption as due to a decelerating shock (they also considered departure from a spherical symmetry). While this finding is consistent with a two-zone picture of a dense equatorial torus and near-empty polar regions, it is likely to be an oversimplification. The situation more likely involves a continuous gradient of density going from equatorial to polar directions. Simulations by \cite{booth2016} further suggests that clumps in the circum-binary regions are likely. In any case, it is highly probable for systems like RS~Oph to have some dense regions (with shocked plasma in ionization equilibrium) and some low density regions where shocked plasma remains out of ionization equilibrium in the early stages of a nova.
\par
Additional asymmetries due to the finite distance between the WD and RG are likely to have a negligible effect on the observed X-ray emission. A spherically symmetric wind from the RG with a mass loss rate of 10$^{-7}$ M$_\odot$ yr$^{-1}$ and a velocity of 10~km~s$^{-1}$ would have a density of 2.2$\times 10^9$ cm$^{-3}$ at a distance (from the center of the RG) of 1~AU.  Its density goes as $r^{-2}$, where $r$ is the distance from the center of the RG.
Ejecta with a velocity of 5000 km~s$^{-1}$ would travel 3~AU per day, compared to the binary separation of $\sim$1 AU. Therefore, after several days, the fact that the nova ejecta originates from the WD, not the RG, leads to only a small perturbation of a simple, spherically symmetric picture. The wind density encountered by the ejecta on Day 5 would be about $9 \times 10^7$ cm$^{-3}$ in this example. Such a wind is sufficiently dense to reach ionization equilibrium (which requires $nt \sim 10^{12}$ s~cm$^{-3}$; \citealt{smith2010}) on a timescale much smaller than the age of the nova eruption.   

\subsection{Variable, partial covering absorption}
 
In the early days of the eruption, the pre-existing wind nebula -- including the equatorial density enhancement \citep{booth2016,ribeiro2009} -- provided a significant source of absorption. This absorbing material had density structures on the size scale of the X-ray emitting region, as suggested by the spectral fits with a covering fraction less than 1. Another potential source of intrinsic absorption was the unshocked portion of the ejecta or the portion of the ejecta was shocked but underwent a catastropic cooling \citep{derdzinski2017}.
Throughout the eruption, the X-ray emission emanated from the shocked shell surrounding unshocked ejecta interior to it. In such a situation, X-rays from the near side of the shell are unabsorbed by the interior matter, while those from the far side are absorbed. The evolution of the column density and covering fraction of the absorber constrain the density structure of the circum-binary material and unshocked ejecta. From Fig. \ref{evol}(d), the column density of the absorption decreased with time after its peak of almost $10^{23}$ cm$^{-2}$ on Day 6. This is consistent with the early time X-ray emitting shock moving through the dense circum-binary ring. The density of the equatorial enhancement remaining in front of the shock decreases with time as the shock propagates into the lower density regions. In addition, the covering fraction of the absorber declined from 1.0 (absorber fully covering the source) at the start of the eruption to 0.4 on Day 16 (Fig. ~\ref{evol}e), indicating strong changes in the circum-binary absorbing material and/or the X-ray emission region itself. In the early days of the eruption, the RG wind outside the nova ejecta contributed the bulk of the intrinsic absorption. After the first few NICER observations, the X-ray emitting region had a significant size compared to that of the binary; our lines of sight to different parts of the X-ray emitting plasma passed through different amount of RG wind. By around Day 16, the nova ejecta had expanded to such a degree that the wind nebula outside the ejecta likely provided little absorption. Instead, it may have been the unshocked nova ejecta (surrounded by the shocked shell) that provided the intrinsic absorption. The front side of the shocked shell was seen only through interstellar absorption, while the back side could have also been absorbed by the unshocked ejecta, resulting in a partial covering fraction close to, but less than 0.4, as we observed around Day 16. The rise in covering fraction back to 1 by around Day 18 is consistent with the drop in absorbing column at around that same time to the interstellar value.

\section{Summary}

\begin{itemize}
\item The 2021 eruption of the recurrent nova RS~Oph was monitored extensively by NICER. Neither single nor two-temperature collisionally-ionized plasma (APEC) model with variable abundances and velocity- and thermally-broadened emission lines could adequately model the line emission from the K shells of medium Z elements such as Ne, Mg, and Si. The APEC models also led to large residuals below approximately 1~keV. We therefore used an alternative approach to modelling the NICER spectra by using a grid of non-equilibrium ionization collisional plasma components with a single temperature and four density-weighted ionization timescales, along with a partial covering absorber. 

\item The evolution of the normalizations of the spectral components with different density-weighted ionization timescales can be understood if the X-ray emission during approximately the first two weeks of the eruption was dominated by shocks propagating through circum-binary material concentrated in the equatorial plane. Because the timescale required for shocked gas to achieve ionization equilibrium is inversely proportional to the square of the gas density, dense shocked regions reach ionization equilibrium more quickly than less dense regions, where the shocked gas remains out of equilibrium for longer. During the first two weeks, emission from plasma close to ionization equilibrium, where the high density-weighted ionization timescales of $10^{11}$ to $10^{12}$ s~cm$^{-3}$ was dominant. After Day 18, the interaction between the ejecta and the less dense circum-binary material in the polar 
regions became more evident -- the contribution to the EM from plasma with the higher density-weighted ionization timescales of $10^{11} - 10^{12}$ s~cm$^{-3}$ decreased, and the contribution from plasma out of ionization equilibrium, with lower density-weighted ionization timescales of $10^9 - 10^{10}$~s~cm$^{-3}$ became important. After Day 6, the EM monotonically decreased as a power-law with index -1.2$\pm$0.1. 

\item After peaking at around $10^{23}$ cm$^{-2}$, the column density absorbing the X-rays followed a decline. This behavior is consistent with the X-ray emission initially being dominated by shocks related to the ejecta interacting with the dense circum-binary ring around the binary and then shocks associated with expansion into the lower density polar regions. The partial covering absorption fraction decreased from near unity at the start of the eruption to 0.4 on Day 16, as expected if absorption around Day 16 was primarily due to unshocked ejecta. 

\item Although some ambiguity is introduced into the interpretation of the temperature evolution by our use of a single temperature to describe X-ray emission from behind multiple shocks, we move forward with such an interpretation assuming that one shock dominated the X-ray emission at any given time. The evolution of the temperature of the non-equilibrium plasma, $kT$, shows an initial increase through Day 5, with a peak value of 24 keV (Fig. \ref{evol}a). The peak in the temperature around Day 4--7 is roughly coincident with the enhancement in X-ray intensity in the MAXI 10--20~keV GSC lightcurve (Fig. \ref{maxi_lc}). The initial rise in the X-ray temperature could have been associated with a decrease in the shock power going into particle acceleration, which can lead to rapid cooling of the post-shock plasma. The detection of RS~Oph in GeV and TeV gamma rays shows that particle acceleration was present in the early days of the 2021 eruption. After Day 5, the temperature decreased as a power-law with an index of -0.9$\pm$0.1 as the eruption evolves, close to the expected decline rate for a shock heated by a decelerating blast wave moving into a stellar wind \citep{sokoloski2006}.

\item The sudden increase in the EM around Day 6 was possibly due to an interaction between the nova ejecta and a clump of matter previously ejected from the RG. We speculate that an episode of enhanced mass loss from the RG about 5 years prior to the nova could have produced such a clump while simultaneously explaining the optical brightening observed in the AAVSO lightcurves during 2017.

\end{itemize}

\section*{Acknowledgement}
We thank the anonymous referee for constructive comments which helped improve the manuscript. We thank Marina Orio and Juan Luna for helpful discussions. The scientific results reported here are based on observations made by the NICER X-ray observatory and we thank the NICER team for scheduling and the execution of these Target of Opportunity observations. This research has made use of MAXI data provided by RIKEN, JAXA and the MAXI team. This research has made use of data and/or software provided by the High Energy Astrophysics Science Archive Research Center (HEASARC), which is a service of the Astrophysics Science Division at NASA/GSFC. J.L.S. acknowledges support from NASA grant 80NSSC21K0715 and NSF award AST-1816100. NI acknowledges support from NASA grant 80NSSC21K1994.

\bibliography{bibtex}
\newpage
\begin{sidewaystable}
\footnotesize
\centering

\caption{Summary of NICER observations} 
\begin{tabular}{c c c c c}
\hline
ObsID  & Start Time  & End Time & Total Exp (ks) & Avg count-rate \\
\hline
4202300101 & 2021-08-10T11:55:17 & 2021-08-10T12:14:20 & 1.02 & 59.3 \\
4202300102 & 2021-08-11T04:56:00 & 2021-08-11T23:41:35 & 2.62 & 70.5 \\
4202300103 & 2021-08-12T01:08:00 & 2021-08-12T22:55:55 & 2.63 & 98.9 \\
4202300104 & 2021-08-13T03:27:10 & 2021-08-13T23:42:22 & 3.88 & 134.6 \\
4202300105 & 2021-08-14T01:09:01 & 2021-08-14T22:59:20 & 3.97 & 154.8 \\
4202300106 & 2021-08-15T03:29:31 & 2021-08-15T23:46:19 & 6.16 & 156.5 \\
4202300107 & 2021-08-16T01:11:22 & 2021-08-16T22:58:13 & 3.85 & 157.4 \\
4202300108 & 2021-08-17T01:59:02 & 2021-08-17T23:47:19 & 3.84 & 154.4 \\
4202300109 & 2021-08-18T01:13:09 & 2021-08-18T23:00:46 & 2.98 & 151.7 \\
4202300110 & 2021-08-19T20:38:03 & 2021-08-19T22:26:11 & 1.04 & 145.2 \\
4202300111 & 2021-08-20T21:26:19 & 2021-08-20T23:15:38 & 1.27 & 140.6 \\
4202300112 & 2021-08-21T00:31:00 & 2021-08-21T00:48:38 & 0.92 & 140.7 \\
4202300113 & 2021-08-21T23:47:07 & 2021-08-22T09:39:08 & 9.61 & 141.3 \\
4202300114 & 2021-08-23T16:05:56 & 2021-08-23T17:41:20 & 0.41 & 141.1 \\
4202300115 & 2021-08-24T15:21:31 & 2021-08-24T15:54:06 & 1.36 & 144.0 \\
4202300116 & 2021-08-25T16:14:40 & 2021-08-25T16:49:00 & 1.41 & 147.4 \\
4202300117 & 2021-08-26T07:41:00 & 2021-08-26T08:19:20 & 2.18 & 146.2 \\
4202300118 & 2021-08-28T01:46:00 & 2021-08-28T05:17:40 & 1.53 & 145.2 \\
4202300119 & 2021-08-30T00:05:00 & 2021-08-30T10:01:20 & 9.51 & 142.6 \\
4202300120 & 2021-08-31T02:25:17 & 2021-08-31T05:46:34 & 2.45 & 135.2 \\
4202300121 & 2021-09-01T04:47:35 & 2021-09-01T11:13:20 & 3.89 & 143.6 \\
4202300122 & 2021-09-02T00:56:40 & 2021-09-02T23:00:49 & 4.47 & 132.6 \\
4202300123 & 2021-09-03T00:11:33 & 2021-09-03T23:47:00 & 6.69 & 140.6 \\
4202300124 & 2021-09-04T07:12:00 & 2021-09-04T21:31:40 & 2.62 & 149.1 \\
4202300125 & 2021-09-05T01:56:40 & 2021-09-05T23:54:20 & 7.08 & 160.3 \\
4202300126 & 2021-09-06T01:02:33 & 2021-09-06T23:27:00 & 8.84 & 124.2 \\
4202300127 & 2021-09-07T02:17:00 & 2021-09-07T23:52:40 & 7.73 & 126.4 \\
4202300128 & 2021-09-08T02:36:03 & 2021-09-08T23:19:20 & 8.33 & 196.7 \\
4202300129 & 2021-09-09T02:17:20 & 2021-09-09T23:53:00 & 8.59 & 227.4 \\
4202300130 & 2021-09-10T02:41:20 & 2021-09-10T23:26:40 & 9.75 & 303.3 \\
4202300131 & 2021-09-11T02:16:40 & 2021-09-11T22:35:20 & 6.73 & 148.7 \\
4202300132 & 2021-09-12T03:03:27 & 2021-09-12T12:28:20 & 2.16 & 157.7 \\
4202300133 & 2021-09-15T03:31:21 & 2021-09-15T20:46:40 & 2.03 & 193.7 \\
4202300134 & 2021-09-16T01:13:34 & 2021-09-16T23:12:20 & 4.41 & 304.4 \\
4202300135 & 2021-09-17T03:35:00 & 2021-09-17T14:44:00 & 3.04 & 307.9 \\

\hline

\end{tabular}
\end{sidewaystable}

\newpage
\begin{sidewaystable}
\footnotesize
\centering

\caption{Best fit spectral parameters for the NICER spectra using the model defined in Section 3.2. The errors are 90\% confidence limits. T$_{0}$ is defined as the start of the optical eruption at MJD 59,434.5} 
\begin{tabular}{c c c c c c c c c c c c c c c}
\hline
Days & kT & $N_{\rm {HW}}$ & CvrF & $N_{\rm H}$ & N1 & N2 & N3 & N4 & Fe & Mg & Si & N & Ne & $\chi^{2}$/dof \\
since T$_{0}$ & (keV) & (10$^{22}$ cm$^{-2}$) & & (10$^{22}$ cm$^{-2}$) & s~cm$^{-3}$ & s~cm$^{-3}$ & s~cm$^{-3}$ & s~cm$^{-3}$ & & & & & & \\
\hline
1.9 & 6.4$^{+0.3}_{-0.4}$ & 5.4$^{+0.4}_{-0.3}$ & 0.87$^{+0.03}_{-0.05}$ & 3.5$^{+0.3}_{-0.4}$ & 0.06$\pm$0.05 & 0.04$\pm$0.02 & 0.23$\pm$0.04 & 0.54$^{+0.07}_{-0.06}$ & 1.09$\pm$0.09 & 1.0 & 1.0 & 1.0 & 1.0 & 195.85/124 \\

2.7 & 8.3$^{+0.4}_{-0.2}$ & 5.3$\pm$0.4 & 0.81$^{+0.02}_{-0.03}$ & 2.5$\pm$0.1 & 0.09$^{+0.01}_{-0.02}$ & 0.034$^{+0.009}_{-0.005}$ & 0.17$\pm$0.03 & 0.50$\pm$0.03 & 1.43$\pm$0.08 & 1.0 & 1.0 & 1.0 & 1.0 & 418.60/134 \\

3.5 & 17.1$^{+0.6}_{-2.5}$ & 4.2$^{+0.2}_{-0.1}$ & 0.88$\pm$0.01 & 1.66$\pm$0.09 & 0.03$\pm$0.02 & 0.08$\pm$0.01 & 0.35$\pm$0.03 & 0.44$^{+0.02}_{-0.03}$ & 1.23$\pm$0.07 & 1.0 & 1.0 & 1.0 & 1.0 & 624.98/136 \\

4.6 & 24$^{+1}_{-3}$ & 2.5$\pm$0.1 & 0.88$\pm$0.2 & 1.12$\pm$0.07 & 0.05$\pm$0.02 & 0.059$\pm$0.01 & 0.35$\pm$0.02 & 0.45$\pm$0.03 & 1.16$\pm$0.07 & 1.0 & 1.0 & 1.0 & 1.0 & 736.36/144 \\

5.5 & 6.9$\pm$0.2 & 8.8$^{+0.7}_{-0.6}$ & 0.71$\pm$0.1 & 1.36$\pm$0.03 & 0.19$\pm$0.03 & 0.04$\pm$0.01 & 0.35$\pm$0.03 & 0.82$^{+0.06}_{-0.04}$ & 0.78$\pm$0.05 & 1.0 & 1.0 & 1.0 & 1.0 & 533.59/144 \\

6.6 & 6.0$\pm$0.2 & 7.8$^{+0.6}_{-0.5}$ & 0.74$\pm$0.01 & 1.10$\pm$0.04 & 0.12$\pm$0.03 & 0.041$^{+0.007}_{-0.008}$ & 0.29$\pm$0.03 & 0.85$^{+0.06}_{-0.07}$ & 0.72$\pm$0.05 & 1.5$\pm$0.2 & 1.6$\pm$0.2 & 1.0 & 1.0 & 344.11/148 \\

7.6 & 5.1$\pm$0.2 & 6.7$^{+0.6}_{-0.5}$ & 0.70$\pm$0.02 & 1.05$\pm$0.04 & 0.06$\pm$0.03 & 0.06$\pm$0.01 & 0.29$^{+0.04}_{-0.03}$ & 0.76$^{+0.08}_{-0.07}$ & 0.63$\pm$0.05 & 1.2$\pm$0.1 & 1.2$\pm$0.1 & 1.0 & 1.0 & 229.33/142 \\

8.6 & 4.4$\pm$0.2 & 5.7$\pm$0.5 & 0.69$\pm$0.02 & 0.99$\pm$0.04 & 0.06$\pm$0.03 & 0.06$\pm$0.01 & 0.28$\pm$0.03 & 0.66$^{+0.07}_{-0.06}$ & 0.61$\pm$0.05 & 1.2$\pm$0.1 & 1.1$\pm$0.1 & 1.0 & 1.0 & 279.24/141 \\

10.4 & 3.9$\pm$0.1 & 4.7$^{+0.6}_{-0.4}$ & 0.66$^{+0.03}_{-0.02}$ & 0.98$\pm$0.04 & 0.05$\pm$0.03 & 0.07$\pm$0.01 & 0.25$\pm$0.03 & 0.59$^{+0.07}_{-0.06}$ & 0.59$\pm$0.05 & 1.1$\pm$0.1 & 0.9$\pm$0.1 & 1.0 & 1.0 & 258.08/138 \\

11.4 & 3.3$\pm$0.2 & 3.7$\pm$0.5 & 0.64$\pm$0.04 & 0.99$\pm$0.04 & 0.03$^{+0.04}_{-0.03}$ & 0.08$\pm$0.01 & 0.22$\pm$0.03 & 0.53$\pm$0.07 & 0.58$^{+0.08}_{-0.07}$ & 1.1$\pm$0.1 & 0.9$\pm$0.1 & 1.0 & 1.0 & 247.30/123 \\

12.3 & 2.9$\pm$0.1 & 3.4$^{+0.5}_{-0.4}$ & 0.63$\pm$0.04 & 1.02$\pm$0.04 & 0.06$\pm$0.04 & 0.08$\pm$0.01 & 0.22$\pm$0.03 & 0.45$\pm$0.06 & 0.59$^{+0.09}_{-0.08}$ & 1.1$\pm$0.1 & 0.8$\pm$0.1 & 1.0 & 1.0 & 273.20/126 \\

12.5 & 2.9$\pm$0.1 & 3.4$^{+0.6}_{-0.5}$ & 0.62$\pm$0.05 & 0.99$\pm$0.05 & 0.04$^{+0.04}_{-0.03}$ & 0.08$\pm$0.01 & 0.19$\pm$0.03 & 0.47$\pm$0.07 & 0.57$^{+0.1}_{-0.09}$ & 1.1$\pm$0.2 & 0.9$\pm$0.1 & 1.0 & 1.0 & 227.36/122 \\

13.5 & 2.77$^{+0.07}_{-0.08}$ & 2.7$^{+0.4}_{-0.3}$ & 0.53$\pm$0.04 & 0.97$^{+0.04}_{-0.03}$ & 0.06$^{+0.03}_{-0.02}$ & 0.059$^{+0.009}_{-0.007}$ & 0.16$\pm$0.02 & 0.39$\pm$0.04 & 0.64$^{+0.06}_{-0.05}$ & 1.1$\pm$0.1 & 0.94$\pm$0.09 & 1.0 
 & 1.0 & 603.94/145 \\

15.2 & 2.5$\pm$0.2 & 2.7$^{+1.7}_{-0.8}$ & 0.4$\pm$0.1 & 1.08$\pm$0.05 & 0.05$^{+0.07}_{-0.04}$ & 0.09$^{+0.02}_{-0.01}$ & 0.15$\pm$0.03 & 0.30$\pm$0.06 & 0.5$^{+0.2}_{-0.1}$ & 0.9$^{+0.2}_{-0.1}$ & 0.8$\pm$0.1 & 1.0 & 1.0 & 243.71/108 \\

16.1 & 2.27$\pm$0.09 & 2.2$^{+0.7}_{-0.5}$ & 0.4$\pm$0.1 & 1.04$\pm$0.04 & 0.08$^{+0.05}_{-0.04}$ & 0.08$\pm$0.01 & 0.13$\pm$0.02 & 0.27$\pm$0.04 & 0.7$^{+0.2}_{-0.1}$ & 1.1$^{+0.2}_{-0.1}$ & 0.89$^{+0.1}_{-0.08}$ & 1.0 & 1.0 & 378.70/125 \\

17.2 & 1.86$\pm$0.09 & 2.1$^{+0.4}_{-0.3}$ & 0.6$\pm$0.1 & 1.02$\pm$0.04 & 0.17$^{+0.07}_{-0.06}$ & 0.11$^{+0.02}_{-0.01}$ & 0.13$\pm$0.02 & 0.25$\pm$0.04 & 1.0$^{+0.3}_{-0.2}$ & 1.2$\pm$0.2 & 0.90$\pm$0.09 & 1.0 &  1.0 & 410.77/127 \\

17.8 & 1.98$\pm$0.09 & 1.7$\pm$0.3 & 0.90$\pm$0.04 & 0.99$\pm$0.04 & 0.14$^{+0.05}_{-0.04}$ & 0.09$\pm$0.01 & 0.12$\pm$0.02 & 0.19$\pm$0.03 & 1.1$^{+0.3}_{-0.2}$ & 1.2$^{+0.2}_{-0.1}$ & 0.95$\pm$0.09 & 1.0 & 1.0 & 421.09/127 \\

19.6 & 1.79$\pm$0.06 & 1.8$\pm$0.5 & 1.0 & 1.00$\pm$0.04 & 0.16$^{+0.04}_{-0.03}$ & 0.09$\pm$0.01 & 0.09$\pm$0.01 & 0.18$\pm$0.02 & 1.0 & 1.1$\pm$0.1 & 0.9$\pm$0.1 & 11$^{+3}_{-2}$ & 0.9$\pm$0.1 & 453.54/127 \\

21.5 & 1.78$^{+0.03}_{-0.02}$ & 0.79$\pm$0.02 & 1.0 & 0.24 & 0.09$\pm$0.02 & 0.069$\pm$0.005 & 0.065$\pm$0.007 & 0.12$\pm$0.01 & 1.0 & 0.98$\pm$0.08 & 0.92$\pm$0.06 & 20$^{+3}_{-1}$ & 0.8$\pm$0.1 & 581.42/142 \\

22.6 & 1.67$\pm$0.05 & 0.81$\pm$0.03 & 1.0 & 0.24 & 0.12$\pm$0.03 & 0.081$\pm$0.008 & 0.07$\pm$0.01 & 0.09$\pm$0.02 & 1.0 & 0.89$\pm$0.09 & 0.84$^{+0.04}_{-0.07}$ & 10$^{+2}_{-1}$ & 0.6$\pm$0.1 & 474.39/127 \\

23.7 & 1.59$\pm$0.05 & 0.78$\pm$0.03 & 1.0 & 0.24 & 0.10$\pm$0.02 & 0.072$\pm$0.006 & 0.049$^{+0.008}_{-0.007}$ & 0.11$\pm$0.01 & 1.0 & 1.12$\pm$0.09 & 0.99$\pm$0.07 & 9$\pm$1 & 1.0 & 456.94/136 \\

24.5 & 1.60$\pm$0.03 & 0.72$\pm$0.02 & 1.0 & 0.24 & 0.07$\pm$0.01 & 0.065$\pm$0.005 & 0.044$^{+0.006}_{-0.005}$ & 0.11$\pm$0.01 & 1.0 & 1.19$\pm$0.07 & 1.0 & 10$\pm$1 & 1.0 & 536.08/130 \\

25.5 & 1.48$\pm$0.03 & 0.71$\pm$0.02 & 1.0 & 0.24 & 0.08$^{+0.02}_{-0.01}$ & 0.063$\pm$0.005 & 0.041$^{+0.007}_{-0.005}$ & 0.094$^{+0.007}_{-0.01}$ & 1.0 & 1.22$^{+0.06}_{-0.05}$ & 1.0 & 20$\pm$2 & 1.0 & 610.73/133 \\

26.8 & 1.39$\pm$0.04 & 0.74$\pm$0.03 & 1.0 & 0.24 & 0.09$\pm$0.02 & 0.067$\pm$0.007 & 0.044$\pm$0.007 & 0.07$\pm$0.01 & 1.0 & 1.17$^{+0.08}_{-0.07}$ & 1.0 & 30$\pm$4 & 1.0 & 594.59/132 \\

27.6 & 1.14$^{+0.03}_{-0.02}$ & 0.77$\pm$0.02 & 1.0 & 0.24 & 0.18$^{+0.02}_{-0.03}$ & 0.059$\pm$0.005 & 0.07$\pm$0.01 & 0.047$\pm$0.006 & 1.0 & 1.01$^{+0.08}_{-0.06}$ & 1.0 & 37$^{+4}_{-3}$ & 1.0 & 703.18/140 \\

28.5 & 1.29$\pm$0.03 & 0.68$\pm$0.02 & 1.0 & 0.24 & 0.09$\pm$0.02 & 0.061$^{+0.005}_{-0.003}$ & 0.029$^{+0.006}_{-0.007}$ & 0.08$\pm$0.01 & 1.0 & 1.27$^{+0.06}_{-0.07}$ & 1.0 & 15$\pm$2 & 1.0 & 558.17/134 \\

29.6 & 1.21$\pm$0.03 & 0.68$\pm$0.02 & 1.0 & 0.24 & 0.10$\pm$0.02 & 0.059$^{+0.006}_{-0.005}$ & 0.026$^{+0.007}_{-0.009}$ & 0.07$\pm$0.01 & 1.0 & 1.30$^{+0.09}_{-0.08}$ & 1.0 & 23$\pm$3 & 1.0 & 605.34/133 \\

30.6 & 0.83$\pm$0.01 & 0.68$^{+0.03}_{-0.02}$ & 1.0 & 0.24 & 0.15$^{+0.04}_{-0.03}$ & 0.023$\pm$0.005 & 0.03$\pm$0.01 & 0.114$\pm$0.004 & 1.0 & 1.25$^{+0.08}_{-0.09}$ & 1.0 & 105$^{+10}_{-19}$ & 1.0 & 568.5/134 \\

31.6 & 0.99$\pm$0.01 & 0.66$\pm$0.02 & 1.0 & 0.24 & 0.13$^{+0.01}_{-0.02}$ & 0.040$\pm$0.005 & 0.03$\pm$0.01 & 0.068$^{+0.002}_{-0.004}$ & 1.0 & 1.32$^{+0.08}_{-0.06}$ & 1.0 & 63$^{+10}_{-8}$ & 1.0 & 619.63/127 \\

32.6 & 1.24$\pm$0.03 & - & - & 0.24 & 0.09$\pm$0.02 & 0.010$^{+0.01}_{-0.009}$ & 0.03$\pm$0.01 & 0.039$\pm$0.009 & 1.0 & 1.0 & 1.0 & 1.0 & 1.0 & 116.64/104 \\

33.6 & 1.19$\pm$0.03 & - & - & 0.24 & 0.09$\pm$0.02 & 0.02$\pm$0.01 & 0.02$\pm$0.01 & 0.05$\pm$0.01 & 1.0 & 1.0 & 1.0 & 1.0 & 1.0 & 71.92/101 \\

34.6 & 1.12$\pm$0.03 & - & - & 0.24 & 0.11$\pm$0.02 & 0.01$\pm$0.01 & 0.02$\pm$0.01 & 0.04$\pm$0.01 & 1.0 & 1.0 & 1.0 & 1.0 & 1.0 & 81.14/90 \\

37.7 & 1.03$\pm$0.06 & - & - & 0.24 & 0.09$\pm$0.03 & 0.02$^{+0.01}_{-0.02}$ & 0.008$^{+0.01}_{-0.008}$ & 0.04$\pm$0.01 & 1.0 & 1.0 & 1.0 & 1.0 & 1.0 & 87.74/98 \\

38.6 & 1.01$\pm$0.04 & - & - & 0.24 & 0.09$^{+0.02}_{-0.01}$ & 0.02$\pm$0.01 & 0.009$^{+0.01}_{-0.009}$ & 0.04$\pm$0.01 & 1.0 & 1.0 & 1.0 & 1.0 & 1.0 & 103.56/96 \\

39.7 & 0.96$\pm$0.03 & - & - & 0.24 & 0.09$^{+0.03}_{-0.02}$ & 0.03$^{+0.01}_{-0.02}$ & 0.004$^{+0.01}_{-0.004}$ & 0.041$^{+0.009}_{-0.006}$ & 1.0 & 1.0 & 1.0 & 1.0 & 1.0 & 82.77/95 \\

\hline

\end{tabular}
\end{sidewaystable}

\end{document}